\documentclass[twocolumn,dvipsnames,prd,nofootinbib,floatfix,superscriptaddress]{revtex4}
\usepackage{xcolor}
\usepackage{amsmath,amssymb}
\usepackage{appendix}
\usepackage{subfigure}
\usepackage{natbib}
\usepackage{todonotes}
\usepackage{graphicx}
\usepackage{hyperref}
\usepackage{footnote}

\hypersetup{
    colorlinks=true,
    linkcolor=black,
    filecolor=magenta,      
    urlcolor=blue,
    pdftitle={Overleaf Example},
    pdfpagemode=FullScreen,
    }

\usepackage{dcolumn}
\usepackage{bm}
\usepackage{orcidlink}
\usepackage{hyperref}
\usepackage{xcolor}
\usepackage{soul}
\sethlcolor{yellow}

\begin{document}

\title{Phase Space of Binary Black Holes from Gravitational Wave Observations to Unveil its Formation History}
\author{Samsuzzaman Afroz \orcidlink{0009-0004-4459-2981}}\email{samsuzzaman.afroz@tifr.res.in}
\author{Suvodip Mukherjee \orcidlink{0000-0002-3373-5236}}\email{suvodip@tifr.res.in}
\affiliation{Department of Astronomy and Astrophysics, Tata Institute of Fundamental Research, Mumbai 400005, India}

\begin{abstract}
Gravitational Wave (GW) sources offer a valuable window to the physical processes that govern the formation of binary compact objects (BCOs). However, deciphering such information from GW data is substantially challenging due to the difficulty in mapping from the space of observation to the space of numerous theoretical models. We introduce the concept of BCO Phase-Space that connects the observable space to the evolution trajectories of the BCO formation channels with cosmic time and apply it to the third GW transient catalog (GWTC-3) that brings new insights into probable astrophysical formation scenarios of nearly $90$ events. Our study reveals that two events, GW190425 and GW230529, show an overlap with a \texttt{BCO Phase Space} trajectory of the same formation channel arising from a sub-solar mass black hole scenario that has grown into a higher mass by accretion, hinting towards the common primordial origin of both these sources. Though the actual formation channel is yet to be confirmed, with the availability of more GW events, the \texttt{BCO Phase Space} can delve into distinguishing features of different formation channels for both astrophysical and primordial origin and opens the possibility of bringing new and deeper insights on the formation and evolution of BCOs across all observable masses over most of the cosmic time.
\end{abstract}

\maketitle

\section{Introduction}

The study of gravitational wave (GW) events has opened a transformative window into our understanding of the universe, allowing us to probe the origins and evolution of binary compact objects (BCOs) \citep{Bailes:2021tot, Arimoto:2021cwc}. These BCOs encode crucial information about their formation processes and evolutionary histories, making it essential to extract and analyze this information to enhance our understanding of binary evolution \citep{Mapelli:2021taw,Iorio:2022sgz,Barrett:2017fcw,Dominik:2012kk,Bailyn:1997xt}. With the availability of a few tens of high signal-to-noise ratio GW events \citep{LIGOScientific:2018mvr,LIGOScientific:2016aoc,KAGRA:2021duu,Bouffanais:2021wcr,Franciolini:2021tla,Cheng:2023ddt,Antonelli:2023gpu,2023ApJ...950..181W,Tiwari:2020otp,Tiwari:2021yvr}, the traditional methods for inferring the population of the GW sources, which is primarily by combining the posteriors of different samples and has been successful in shedding light on the mass distribution and merger rate distribution of the binary compact objects. But the main physics question remains unknown, that is what are the different formation channels though binary compact objects form, and how do they evolve over the cosmic time?

\begin{figure*}
\centering
\includegraphics[width=1\textwidth, height=14.0cm]{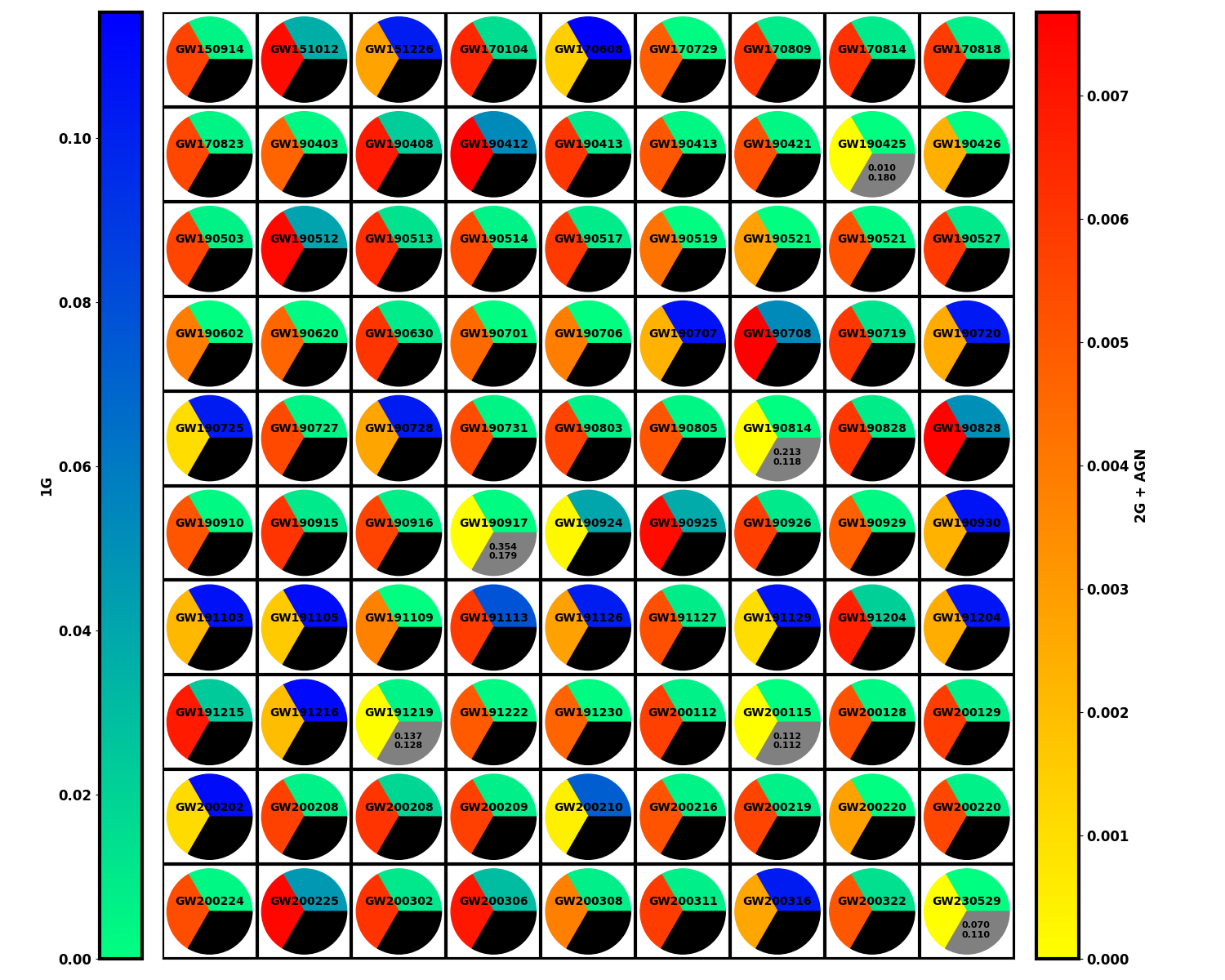}
\caption{This figure displays the probabilities of each compact binary event from GWTC-3 and GW230529 (excluding GW170817, which is confirmed to be a BNS system \citep{LIGOScientific:2017vwq}) originating from either 1G BBH or 2G BBH combined with BBH in AGN disks (2G+AGN) formation channels. Each pie chart represents the relative probabilities of the event being of 1G (top right side) and 2G+AGN (left side) origin. For events with significant probability contributions from either 1G or 2G+AGN, the lower portion of the pie is shown in black, while for events with negligible probability contributions, it is shown in gray for the \textit{confusing sources}. The overlaid numbers for these events represent the $\mathrm{M_{int}}$ (top) and $\mathrm{\dot{m}}$ (below) values for which the event could be considered a candidate for PBH. Note that these values represent only one possible combination; many other combinations are also possible. These probabilities represent relative probabilities for a particular formation channel across different events. These values are not expected to sum to unity for any individual event, as they are not normalized per event. The event-level probabilities should be interpreted within that channel-specific context.}
\label{fig:GWCatelog}
\end{figure*}

To answer the above-mentioned astrophysical question, we need to understand the mapping between the space of the formation channels of compact objects with the space of GW observations. The mapping between these two spaces can connect the formation and evolution history of binaries with the correlations between observable quantities in the GW data. In this work, we propose a new approach the \texttt{BCO Phase Space} which can capture the evolution track of different formation scenarios of BCOs in terms of the observable quantities of the GW sources such as masses, spins, luminosity distance, eccentricity, kick velocity, or any other observable \footnote{We do not consider eccentricity and kick velocity in the remaining analysis as it is not well measured from current GW data.}. The observed GW events capture different regions in the \texttt{BCO Phase Space} and its overlap with the trajectories may be able to identify the possible formation channel. Each formation channel whether resulting from isolated binary evolution, dynamical interactions in dense star clusters, or primordial origins leads to distinct trajectories in this phase space. More importantly, \texttt{BCO Phase Space} enables us to identify not only known formation channels, but also any new population of the GW sources that can exist in nature, but not predicted by any simulations. This method provides a more nuanced perspective on the underlying processes both of astrophysical and primordial origin that influence the formation and evolution of binary systems.

We demonstrate this new \texttt{BCO Phase Space} technique on the 90 events from GWTC-3 \citep{KAGRA:2021duu} along with GW230529 \citep{LIGOScientific:2024elc} from the fourth observation run detected by the LIGO\citep{LIGOScientific:2016dsl}-Virgo\citep{VIRGO:2014yos}-KAGRA\citep{KAGRA:2020tym} (LVK) collaboration to get an insight on the formation channel of the compacts objects with the possibility for both astrophysical and primordial origin. We identify possible overlap of these events with the astrophysical formation channel scenarios arising from the first generation, second generation, and mergers in AGN disc for these binary black holes (BBHs) demonstrating what are their possibilities of overlapping with different formation channels. Furthermore, we classify these events based on their masses and assign probabilities indicating their likely formation channels, as summarized in Figure~\ref{fig:GWCatelog}. The plot shows the probability of these events appearing from different formation scenarios which are considered in this paper. More details on the analysis behind this key result of the paper are given in section \ref{sec:classification}. Due to only approximately 90 events from GWTC-3 available for population analysis, we only consider a limited number of formation channels in this paper to demonstrate the applicability of \texttt{BCO Phase Space} on current data. In the future, with the availability of more GW data, one can consider any number of models in phase space.  

In this work, we only focus on the BBHs scenario and do not consider the neutron star. The \texttt{BCO Phase Space} of a neutron star will exhibit very complementary and rich information about the system. So we plan to consider it in a separate work. We consider all the low mass events except GW170817 \citep{LIGOScientific:2017vwq} in the analysis for finding its overlap with the \texttt{BCO Phase Space} of BBH origin (for both astrophysical and primordial scenario). This is because the true nature of these sources (black hole or neutron star) cannot be confirmed until independent observation such as electromagnetic counterparts \citep{2012ApJ...746...48M,Troja:2017nqp} or tidal deformation \citep{LIGOScientific:2018cki,Flanagan:2007ix} are made.

This paper is organized as follows. In Section~\ref{sec:PhaseSpaceModel}, we characterize the different formation channels of binary black holes using phase space analysis. In Section~\ref{sec:PhaseSpaceGen}, we reconstruct the phase space for these binary systems. In Section~\ref{sec:classification}, we classify all detected gravitational-wave events from GWTC-3. Section~\ref{sec:Result} presents the framework of the phase space technique and our results. Finally, Section~\ref{sec:summary} summarizes our key findings and outlines future prospects.

\section{Characterizing the Formation Pathways for Black Holes in Phase space} 
\label{sec:PhaseSpaceModel}

Astrophysical black holes (ABHs) and primordial black holes (PBHs) represent two distinct yet complementary populations that contribute to GW observations. While ABHs originate from the collapse of massive stars, PBHs can form in the early universe due to large density fluctuations. Understanding their formation mechanisms, mass distributions, and spin evolution is crucial for distinguishing their contributions to GW signals.

\subsection{Astrophysical Black Holes}

The formation of ABHs is thought to proceed through several key channels. Understanding these formation pathways is crucial for constructing accurate astrophysical models and making reliable theoretical predictions. Broadly, ABHs are produced via (i) isolated binary evolution, (ii) hierarchical mergers, and (iii) dynamical interactions in dense stellar environments. In this work, we focus on the active galactic nucleus (AGN) disk channel as a representative dynamical environment, while noting that the framework is sufficiently flexible to encompass additional channels as observational constraints improve.

\subsubsection{Isolated Binary} 
In the isolated binary evolution channel, black holes are born from the core collapse of massive stars within binary systems, largely free from external perturbations. This pathway predominantly produces first-generation (1G) black holes, whose masses are determined by the underlying stellar evolution processes. In particular, the pair-instability supernova mechanism is believed to impose an upper mass limit of approximately 45-50~\(M_\odot\) on 1G black holes \citep{Farmer:2020xne,Vink:2024dgm}. However, uncertainties remain due to factors such as fallback treatment, angular momentum transfer during a collapse, and the influence of metallicity with lower-metallicity stars losing less mass and hence forming more massive black holes \citep{Zhang:2004kx,Mandel:2018hfr}. In addition, the evolutionary history including episodes of mass transfer or common envelope phases affects both the final mass and spin distributions of the black holes. The merger rate distribution for these systems is often modeled using the Madau-Dickinson star formation rate (SFR) in conjunction with an appropriate delay time distribution \citep{Madau:2014bja,Dominik:2014yma,Karathanasis:2022hrb,Mukherjee:2021rtw,Karathanasis:2022rtr}.

The delay time (\( t_d \)) refers to the elapsed period between the formation of stars that ultimately evolve into black holes and the subsequent merging of these black holes. Importantly, this time delay is not uniform across all BBHs but rather follows a specific distribution that accounts for the variability in delay times. This distribution function is defined as follows\citep{Dominik:2014yma, Karathanasis:2022hrb,Mukherjee:2021rtw,Karathanasis:2022rtr}:
\begin{equation}
    \mathrm{p_t(t_d|t_d^{min},t_d^{max},d) \propto 
    \begin{cases}
    (t_d)^{-d} & \text{, for }  t_d^{min}<t_d<t_d^{max}, \\
    0 & \text{otherwise},
    \end{cases}}
\end{equation}
here, the delay time is given by \( t_d = t_m - t_f \), where \( t_m \) and \( t_f \) represent the lookback times of the merger and formation, respectively.

For the 1G BBH merger rate at a given redshift \( z \), we can express it as:
\begin{equation}
    \mathrm{R_{1G}(z) = R_0 \frac{\int_z^{\infty} p_t(t_d|t_d^{min},t_d^{max},d) R_{SFR}(z_f) \frac{dt}{dz_f} dz_f}{\int_0^{\infty} p_t(t_d|t_d^{min},t_d^{max},d) R_{SFR}(z_f) \frac{dt}{dz_f} dz_f}}.
\end{equation}
In this equation, the parameter \( R_0 \) denotes the local merger rate, which indicates the frequency of mergers at redshift \( z = 0 \). According to the study by \cite{KAGRA:2021duu}, the estimated values of \( R_0 \) for the BBH merger rate range between 17.9 \( \mathrm{Gpc^{-3} \, yr^{-1}} \) and 44 \( \mathrm{Gpc^{-3} \, yr^{-1}} \) at a fiducial redshift of \( z = 0.2 \). In our analysis, we adopt a standard local merger rate of \( R_0 = 20 \, \mathrm{Gpc^{-3} \, yr^{-1}} \) for the BBH system. Here, \( R_{SFR}(z_f) \) denotes the star formation rate \citep{Madau:2014bja}, and \( \frac{dt}{dz_f} \) represents the Jacobian of the transformation from cosmic time to redshift.

\subsubsection{Hierarchical Mergers} 

Hierarchical mergers refer to the process in which black holes formed from earlier mergers undergo further mergers with other black holes, resulting in higher-generation black holes. In this study, we focus primarily on second-generation (2G) black holes, which are direct products of first-generation black hole mergers (1G+1G). These 2G black holes emerge when two 1G black holes merge, allowing them to surpass the mass limits imposed by stellar evolution, particularly the pair-instability mass gap. Unlike black holes formed through stellar collapse, 2G black holes retain around 95\% of the combined mass of their progenitor black holes \citep{Pretorius:2005gq,Ossokine:2017dge}. Additionally, these black holes tend to have higher spins, typically clustering around a spin parameter of 0.7, as a result of the merger dynamics \citep{Scheel:2008rj,Campanelli:2006uy,Fishbach:2017dwv}.

\begin{figure}
\centering
\includegraphics[width=9cm, height=5.8cm]{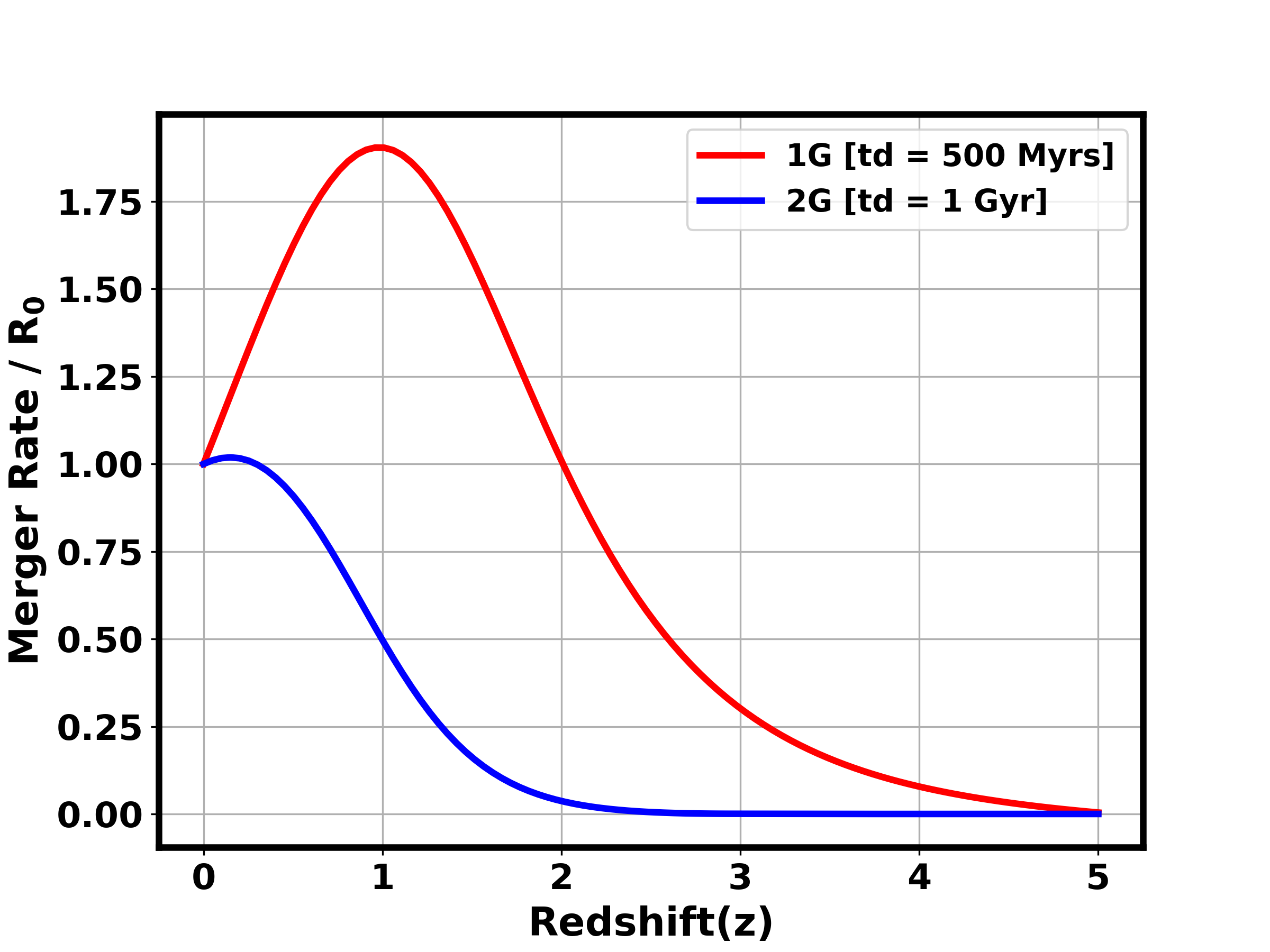}
\caption{The figure shows the delay time merger rates for first-generation (1G) and second-generation (2G) black holes. The delay time for 1G black holes is set at 500 Myrs, while for 2G black holes it is set at 1 Gyr.}
\label{fig:ABHMergRate}
\vspace{-0.5cm}
\end{figure}

For modeling 2G black holes, we assume a hierarchical formation scenario where 2G mergers originate directly from 1G mergers. Thus, the 2G merger rate is determined by convolving the previously calculated 1G merger rate with a delay-time distribution. This convolution naturally incorporates the time delay between successive mergers, causing the overall 2G merger rate to become suppressed and shifted toward lower redshifts, as illustrated in Figure~\ref{fig:ABHMergRate}. While the \texttt{BCO Phase Space} framework is flexible enough to accommodate more detailed and complex merger scenarios, we employ a simplified delay-time model for this initial application:

For modeling 2G black hole mergers, we assume a hierarchical formation scenario where 2G mergers originate directly from 1G mergers. Consequently, the 2G merger rate is determined by convolving the previously calculated 1G merger rate with a delay-time distribution. This convolution naturally accounts for the time delay between successive mergers, leading to a suppression and a shift in the overall 2G merger rate toward lower redshifts, as illustrated in Figure \ref{fig:ABHMergRate}. Although the \texttt{BCO Phase Space} framework can flexibly accommodate more detailed and complex merger scenarios, we initially adopt a simplified delay-time model:

\begin{equation}
    \mathrm{R_{2G}(z) = R_0 \frac{\int_z^{\infty} p_t(t_d|t_d^{min},t_d^{max},d) R_{1G}(z_f) \frac{dt}{dz_f} dz_f}{\int_0^{\infty} p_t(t_d|t_d^{min},t_d^{max},d) R_{1G}(z_f) \frac{dt}{dz_f} dz_f}}.
\end{equation}

In general, delay times are influenced by cluster parameters such as mass, radius, and escape velocity \citep{Stegmann:2022ruy, Chattopadhyay:2023pil, Antonini:2024het}. However, our simplified approach captures the essential characteristics of the mergers without introducing additional uncertainties arising from detailed cluster modeling. Consequently, the derived 2G merger rate effectively reflects the hierarchical nature of black hole mergers by exhibiting suppression and a shift towards lower redshifts. While this model remains sufficiently general to represent diverse merger scenarios including those significantly affected by cluster dynamics—the current observational data, limited by the small number of high signal-to-noise ratio (SNR) events, cannot yet fully constrain all detailed parameters or more complex merger models. Nevertheless, the \texttt{BCO Phase Space} framework can be readily adapted to accommodate alternative merger rate scenarios when richer observational datasets become available.

In Figure \ref{fig:ABHMergRate}, we depict the delay time merger rates for 1G and 2G black holes. The delay time represents the period between the formation of the black hole binary and their eventual merger. For 1G black holes, the delay time is set at 500 million years (Myrs), as indicated by the corresponding curve. In contrast, 2G black holes exhibit a longer delay time, with mergers occurring around 1 billion years (Gyr). The figure highlights the temporal difference between the two populations, with 2G black holes typically experiencing longer periods before merging compared to their 1G counterparts.

\subsubsection{Dynamical Interactions in Dense Stellar Environments} Dynamical formation channels come into play in environments where high densities foster frequent gravitational encounters among stars and compact objects. In globular clusters, nuclear star clusters, and young stellar clusters, BBHs are typically assembled via dynamical captures and exchange interactions rather than through isolated binary evolution \citep{Miller:2008yw,Banerjee:2010,Samsing:2014}. As a result, the BBHs emerging from these channels often exhibit broader and shifted mass distributions with extended high-mass tails a signature of hierarchical mergers and dynamical mass segregation \citep{Fragione:2023kqv,Rodriguez:2016,Zevin:2021}. Their spin distributions are also distinct, with dynamically assembled binaries tending to display more randomized spin orientations compared to the more aligned spins expected from isolated evolution \citep{Antonini:2024het,Samsing:2018}. In this study, we narrow our focus to the AGN disk channel a subset of dynamical environments due to its relatively well-constrained observational properties. In AGN disks, the high-density, gas-rich medium drives frequent interactions among black holes, leading to enhanced merger rates (for further details, refer to Appendix \ref{sec:AGNMerg}), significant mass growth via accretion and repeated mergers, and a tendency for spins to align with the disk’s angular momentum \citep{Tagawa:2019osr,2024arXiv240216948F,Wang:2021clu,Banerjee:2016ths,Rodriguez:2019huv}. Although AGN disks provide a representative dynamical scenario, our phase-space framework is designed to be general. By considering a flexible parametrization of mass and spin distributions, we can readily accommodate the complex dynamical effects observed in other dense stellar environments. However, due to the current limited number of high signal-to-noise ratio events, the available data cannot yet fully constrain all the detailed astrophysical complexities inherent in the cluster channel or any more complex models. Nonetheless, our framework is inherently adaptable and can incorporate these effects as further observational constraints become available.

\subsubsection{Characterizing Mass and Spin of Astrophysical Origin Black Holes}
\label{sec:MassSpinABH}

We parameterize the mass function of astrophysical-origin black holes as a combination of a Gaussian component and an exponential decay term, allowing it to capture a broader class of mass distributions, as defined below:

\begin{equation}
P(m) = 
    \begin{cases}
        \frac{1}{\sqrt{2\pi \sigma^2}} \exp\left(-\frac{(m - M_{median})^2}{2 \sigma^2}\right), & \\ 
        \qquad \text{if } m < M_{median}, \\[10pt]
        \frac{1}{\sqrt{2\pi \sigma^2}} \exp\left(-\frac{(m - M_{median})^2}{2 \sigma^2}\right) \\
        \qquad \times \exp(-\alpha (m - M_{median})), & \\
        \qquad\text{if } m \geq M_{median}.
    \end{cases}
    \label{eq:MassModel}
\end{equation}

\begin{figure}
\centering
\includegraphics[width=9cm, height=5.8cm]{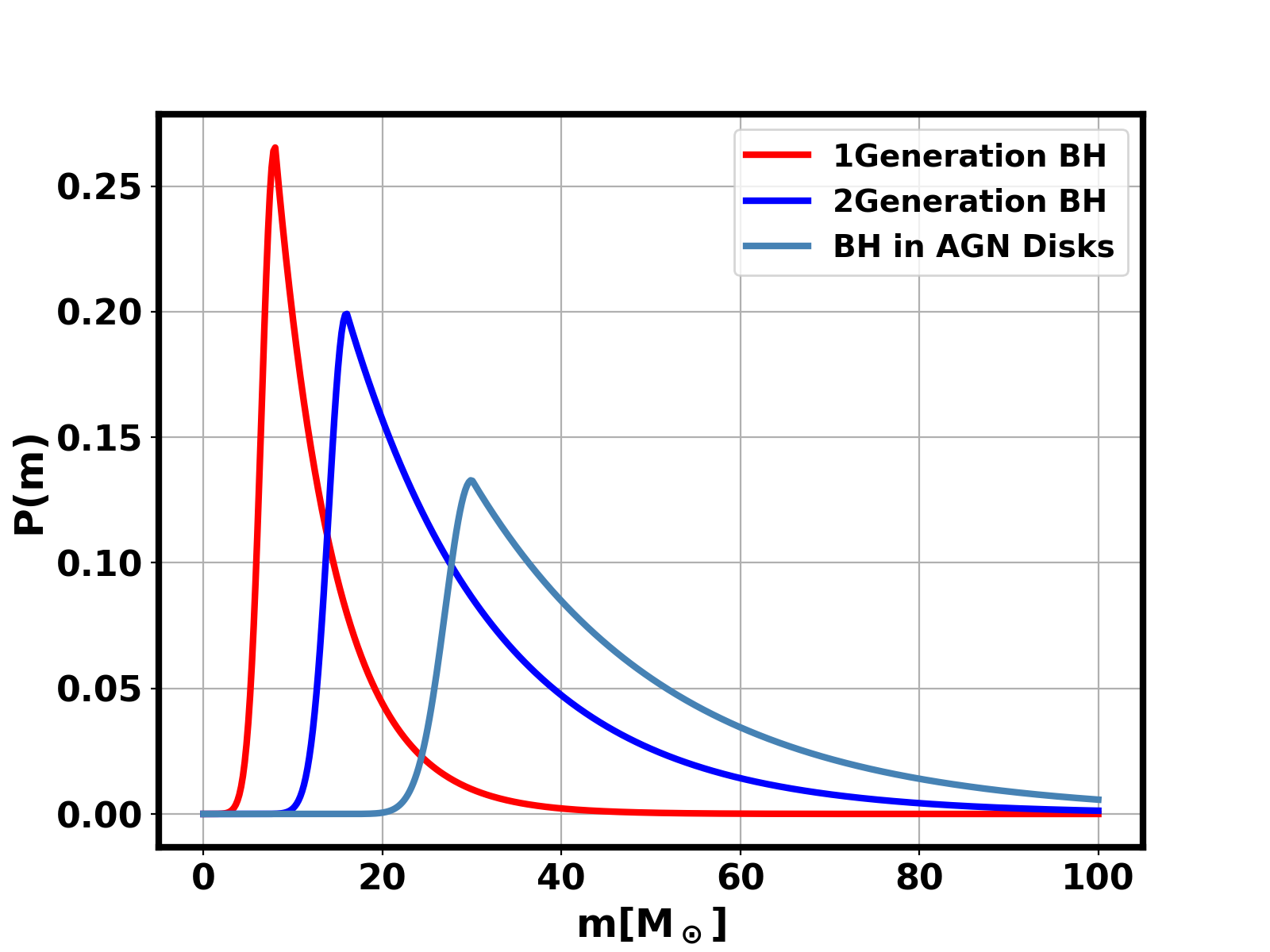}
\caption{The probability density functions for first-generation (1G), second-generation (2G) black holes, and black holes in AGN disks are shown. The 1G black holes (red curve) have a peak at 8 $M_\odot$, while the 2G black holes (blue curve) peak at 16 $M_\odot$. Black holes in AGN disks (steel blue curve) are characterized by a peak at 30 $M_\odot$.}
\label{fig:ABHMassModel}
\end{figure}

\begin{figure}
\centering
\includegraphics[width=9cm, height=5.8cm]{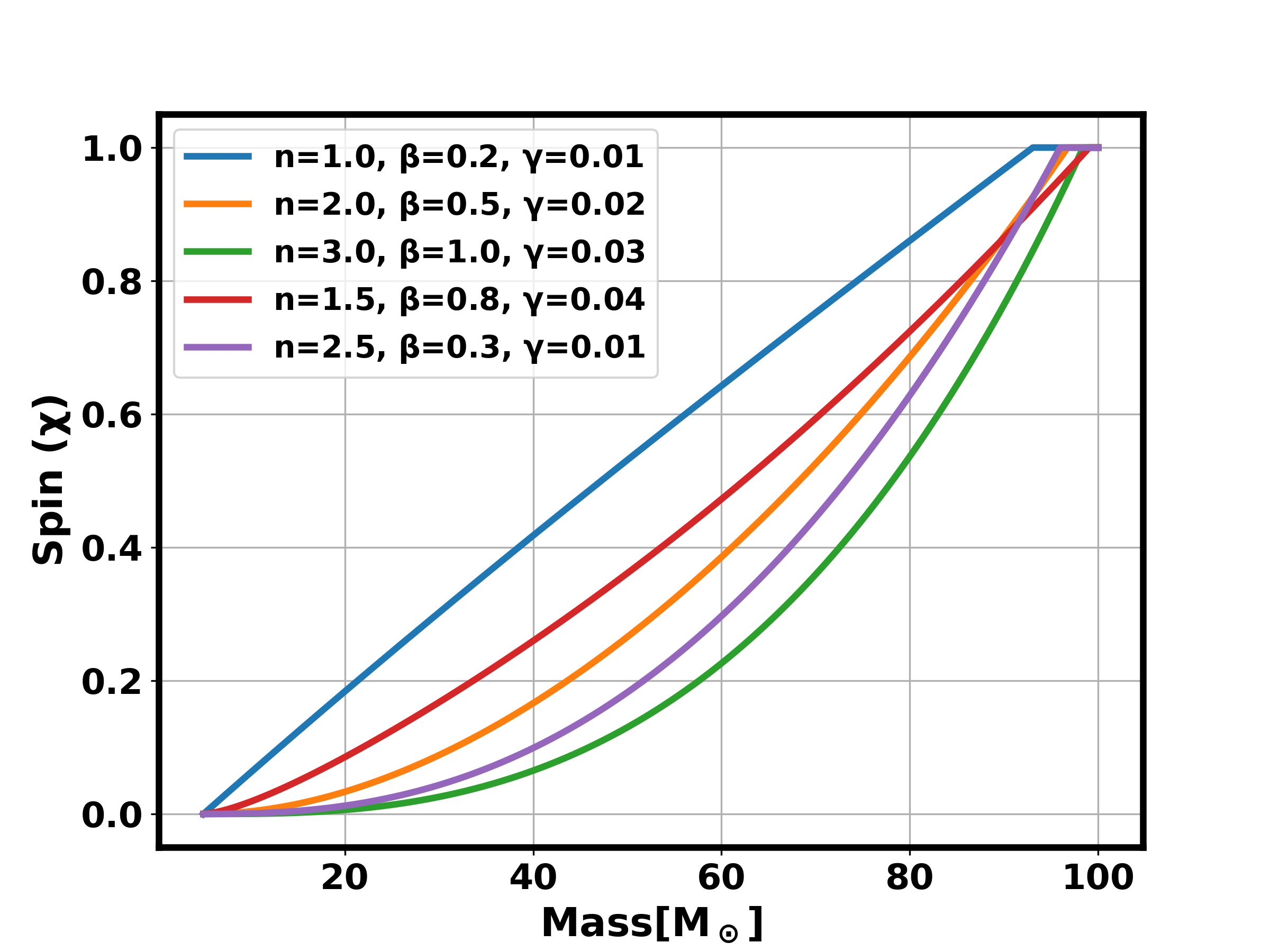}
\caption{Plot of the spin parameter $\mathrm{\chi(m)}$ as a function of mass($\mathrm{m}$) for varying parameter values $\mathrm{n}$, $\mathrm{\beta}$, and $\mathrm{\gamma}$. This illustrates how adjustments in these parameters influence the growth and saturation behavior of the spin parameter in black holes.}
\label{fig:ABHMassCorSpin}
\end{figure}

The combination of Gaussian and exponential terms allows for flexibility in adapting to various astrophysical mass distributions, effectively capturing both central peaks and tails. The inclusion of the exponential decay term also provides the capability to represent asymmetries in mass distributions, which are common in astrophysical observations. Parameters \(M_{median}\), \(\sigma\), and \(\alpha\) enable straightforward adjustments to the peak location, width, and decay rate, facilitating alignment with observed data. Each parameter carries clear physical meaning, enhancing discussions about the processes that shape mass distributions. Moreover, while the model is sufficiently general to encompass the mass distributions expected from cluster channels through appropriate parameter choices, it is important to note that the current dataset comprising only a limited number of GW sources cannot yet fully constrain all aspects of the model, nor can it constrain any more complex models incorporating additional astrophysical details.

In Figure \ref{fig:ABHMassModel}, we illustrate the probability density functions for three distinct black hole populations: 1G, 2G, and black holes formed in AGN disks. The red curve represents the mass distribution of 1G black holes, characterized by a peak at 8 $M_\odot$, indicating that the majority of these black holes are clustered around this mass. The blue curve corresponds to 2G black holes, with a higher peak at 16 $M_\odot$, reflecting their larger mass compared to 1G black holes. Finally, the steel blue curve shows the distribution of black holes in AGN disks, which have an even larger peak at 30 $M_\odot$. These differences in mass distributions highlight the varying characteristics of black holes formed through different evolutionary channels.

In the study of binary systems, particularly black holes, the spin parameter \(\chi_1\) plays a critical role in understanding merger dynamics. We present a generalized model for the spin parameter as a function of mass, incorporating both mass-dependent growth and saturation mechanisms. The equation is given by:

\begin{equation}
\begin{split}
    \chi(m) = &\left( \frac{m - m_{\text{min}}}{m_{\text{max}} - m_{\text{min}}} \right)^n \\ & \times \left[ 1 - \exp\left( -\beta \frac{m - m_{\text{min}}}{m_{\text{max}} - m_{\text{min}}} \right) \right]^\gamma,
\end{split}
\label{eq:massspinabh}
\end{equation}
where \(m_{\text{min}}\) and \(m_{\text{max}}\) define the mass range for spin evolution, \(n\) controls the initial spin growth, and \(\beta\) and \(\gamma\) govern the onset and sharpness of spin saturation at higher masses. This model captures key aspects of black hole spin dynamics, including power-law growth for lower masses and smooth saturation as the mass approaches \(\mathrm{m_{max}}\). By constraining the spin within a realistic range, the model provides flexibility for various astrophysical scenarios, from isolated accretion to hierarchical mergers \citep{Stevenson:2022djs, Zhang:2004kx,Mandel:2018hfr,Barber:2023hvo,Miller:2008yw,Fragione:2023kqv,Britt:2021dtg,Kimball:2020qyd}. It allows for fine-tuning of the spin evolution by adjusting the parameters, ensuring that the spin behavior remains physically accurate and adaptable to observational data. The spin parameter evolves based on the black hole's mass, reflecting both growth and eventual saturation. This framework is valuable for interpreting gravitational wave detections and other astrophysical signals, enabling better constraints on the spin distribution of black holes and their formation pathways.

\subsection{Primordial black holes}
\label{sec:PBHDetails}

Primordial black holes (PBHs) are theorized to have originated from the gravitational collapse of density fluctuations in the early Universe, distinct from astrophysical black holes (ABHs) formed via stellar evolution and mergers. PBHs can exhibit a broad mass range and are considered viable dark matter candidates. Their initial mass distribution, determined at formation, critically influences their subsequent growth through accretion processes, which are highly sensitive to redshift and local environmental conditions.

As PBHs accrete surrounding matter over cosmic time, they not only gain mass but also angular momentum, evolving from nearly non-spinning states toward spins approaching unity. These accretion-driven evolutionary pathways shape their observable characteristics and differentiate their contributions to the overall black hole population.

The merger dynamics of PBHs, including clustering due to gravitational interactions particularly significant during the matter-dominated era, directly impact their detectability as GW sources \citep{Bird:2016dcv,Clesse:2016vqa,Sasaki:2016jop}. Accurate modeling of their merger rates as functions of redshift or luminosity distance is vital for interpreting GW signals observed by current and future detectors like LVK. Studying these mergers can thus provide critical insights into the nature of dark matter and the physics of the early Universe. In our framework, the merger rate as a function of redshift is translated into a rate with respect to luminosity distance. The merger rate of PBHs is given by \citep{Raidal:2018bbj, Vaskonen:2019jpv}

\begin{equation}
\begin{aligned}
    \mathrm{dR_0=\frac{1.6\times10^6}{Gpc^3\,yr}f_{pbh}^{\frac{53}{37}}\eta^{-\frac{34}{37}}\biggl(\frac{M}{ M_\odot}\biggr)^{-\frac{32}{37}}\biggl(\frac{\tau}{t_0}\biggr)^{-\frac{34}{37}}}  \\ \times  \mathrm{0.24\biggl(1+\frac{2.3S_{eq}}{f^2_{pbh}}\biggr)\psi(m_1)\psi(m_2)dm_1dm_2}, 
\end{aligned}
\end{equation}
where $\mathrm{\tau}$ is the time elapsed since the formation of PBHs, $\mathrm{t_0}$ is the current age of the universe, and the term $\mathrm{\biggl(1+\frac{2.3S_{eq}}{f^2_{pbh}}\biggr)}$ reflects the influence of adiabatic perturbations on the eccentricity of early binaries, with $\mathrm{\sqrt{S_{eq}} = 0.0005}$ representing the variance of matter perturbation at matter-radiation equality \citep{Ali-Haimoud:2017rtz,Eroshenko:2016hmn}. The parameter $\mathrm{f_{PBH}}$ denotes the fraction of dark matter composed of PBHs, $\mathrm{\eta}$ is the symmetric mass ratio, and $\mathrm{M}$ is the total mass of the binary. The mass distribution of PBHs, denoted as $\mathrm{\psi(M)}$, is crucial for understanding their formation and abundance in the universe. This distribution is normalized so that integrating over all masses yields the total PBH density, reflecting how PBH populations are spread across different mass ranges \citep{Carr:2017jsz, Horowitz:2016lib}. 

\begin{figure}
\centering
\includegraphics[width=9cm, height=5.8cm]{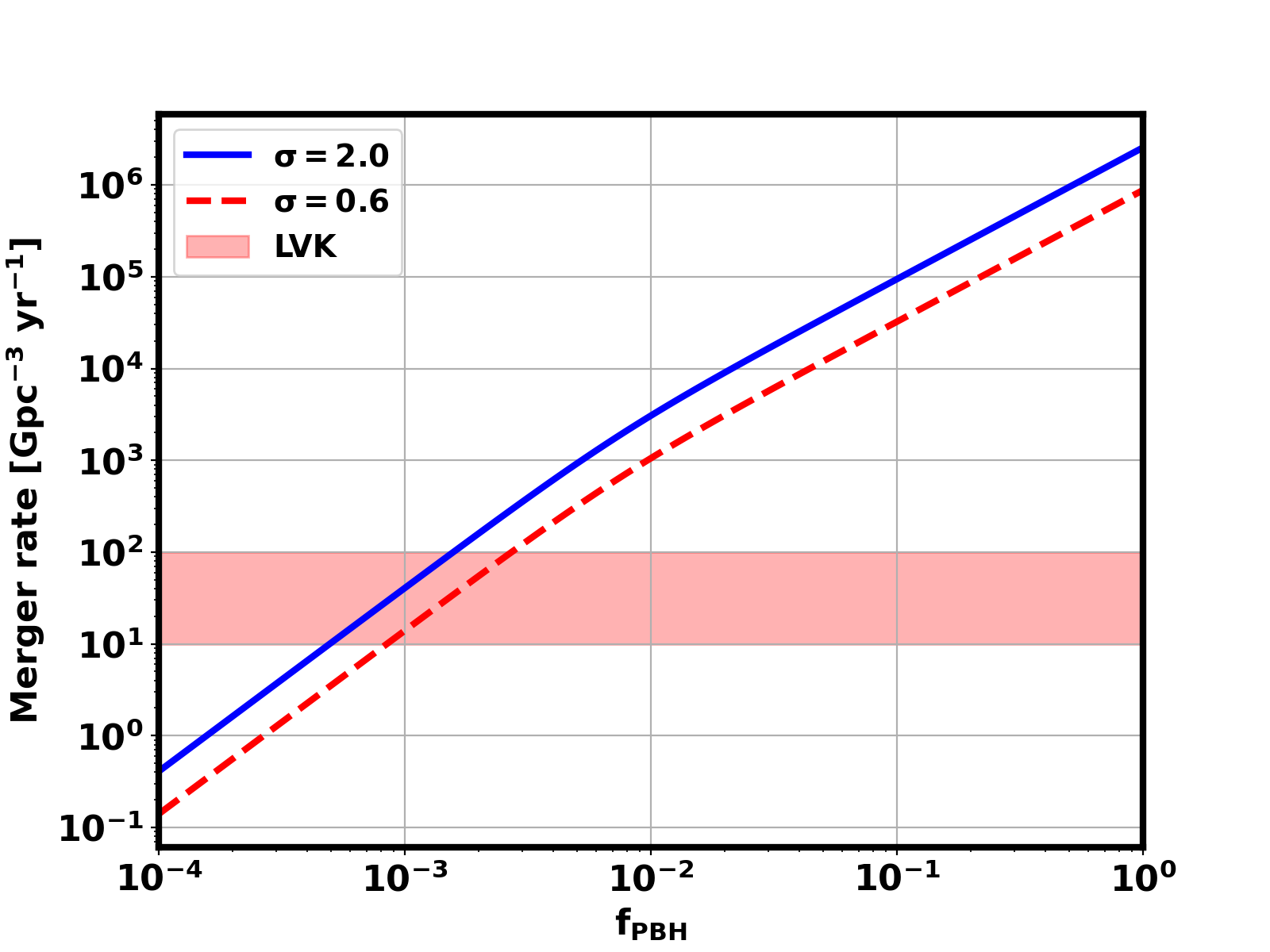}
\caption{Integrated merger rate $\mathrm{R}$ as a function of $\mathrm{f_{PBH}}$ for lognormal mass functions with a characteristic mass $ \mathrm{M_c = 20 \, M_\odot}$. The solid blue line represents the lognormal mass function with a width $\mathrm{\sigma = 2.0}$, while the dashed red line corresponds to a width $\mathrm{\sigma = 0.6}$. The shaded red region indicates the merger rate range consistent with LVK observations.}
\label{fig:PBHMergRateplot}
\end{figure}

PBH formation is linked to the collapse of density perturbations in the early universe, with their masses determined by the scale of these perturbations. Critical collapse results in a broadening of the mass function, allowing PBHs of varying masses to form simultaneously in different regions. Once formed, PBHs behave like non-relativistic matter, and their abundance evolves differently from the total energy density of the universe, which is dominated by radiation in the early stages. A widely used approximation for this distribution is the log-normal distribution \cite{Carr:2018rid}
\begin{equation}
\mathrm{\psi(M) = \frac{1}{\sqrt{2\pi} \sigma M} \exp \left[ - \frac{\ln^2(M/M_c)}{2\sigma^2} \right]},
\end{equation}
where $\mathrm{M_c}$ is the characteristic mass and $\mathrm{\sigma}$ defines the width of the distribution. This parameterization allows for easy adjustments to fit observational constraints and explore a wide range of potential PBH populations \citep{PhysRevD.47.4244,Green:2016xgy,Kannike:2017bxn,Kuhnel:2017pwq}.

In Figure \ref{fig:PBHMergRateplot}, we present the integrated merger rate \( \mathrm{R} \) as a function of the fraction of primordial black holes \( \mathrm{f_{PBH}} \), using lognormal mass functions characterized by a central mass of \( \mathrm{M_c = 20 \, M_\odot} \). The solid blue line illustrates the lognormal mass function with a width of \( \mathrm{\sigma = 2.0} \), while the dashed red line represents a narrower width of \( \mathrm{\sigma = 0.6} \). The variation in the width of the mass function affects the overall merger rate, as seen by the distinct behaviors of the two lines. Additionally, the shaded red region highlights the range of merger rates that are consistent with the observations from the LVK collaboration, providing context for the implications of primordial black hole fraction on detectable merger events.

Depending on the formation scenario \citep{Jedamzik:1996mr,Garcia-Bellido:1996mdl,Crawford:1982yz,Musco:2012au}, the initial mass distribution critically influences the subsequent growth of PBHs through accretion \citep{Mack:2006gz,Serpico:2020ehh}. The maximum accretion rate, dictated by the Eddington limit which balances gravitational attraction against radiation pressure is expressed as \citep{Franciolini:2021nvv}:

\begin{equation}
\mathrm{\dot{M} = \dot{m} \dot{M}_{Edd} = \dot{m} \times 2.2 \, M_{\odot}/\text{Gyrs} \left(\frac{M}{M_{\odot}}\right)},
\label{eq:MassEvoPBH}
\end{equation}
where $\mathrm{\dot{M}_{\text{Edd}}}$ represents the Eddington accretion rate, $\mathrm{M}$ denotes the mass of the PBH, and $\mathrm{\dot{m}}$ is the mass accretion index. For PBHs with very low initial masses on the order of $\mathrm{10^{-5}M_\odot}$ and high mass accretion rates, the final masses can grow to values within the range of current GW detector sensitivity within the age of the universe. Conversely, higher initial masses of the order of 1 $\mathrm{M_\odot}$ with lower accretion rates can also result in final masses that fall within the detectable range of GW observations.

A value of $\mathrm{\dot{m}} = 1$ corresponds to the Eddington accretion rate, while values less than 1 indicate sub-Eddington accretion. Additionally, there is a relationship between mass accretion and spin \citep{DeLuca:2020bjf}. As PBHs accrete matter, their spin increases due to the angular momentum gained from the infalling material. The spin parameter, $\mathrm{\chi}$, evolves from an initial value of zero to a maximum of one, depending on the redshift $\mathrm{z}$ and the mass accretion rate $\mathrm{\dot{m}}$. We parametrize the spin parameter, $\mathrm{\chi}$, as a function of redshift $\mathrm{z}$ and mass accretion rate $\mathrm{\dot{m}}$, given by the following equation:
\begin{equation}
\mathrm{\chi(z, \dot{m}) =  1 - e^{-k \cdot \dot{m} \cdot \Delta t}},
\label{eq:SpinEvoPBH}
\end{equation}
where $\mathrm{\Delta t}$ is the time difference (measured in Gyrs) between the initial redshift $\mathrm{z_{\text{initial}}}$ and the current redshift $\mathrm{z}$, and $\mathrm{k}$ is a constant controlling the rate of spin evolution. Different values of $\mathrm{k}$ lead to varying spin growth rates.

\section{Reconstruction of Phase Space for Binary Black Holes}
\label{sec:PhaseSpaceGen}

To generate the phase space for ABHs and PBHs, we divide the redshift range from $\mathrm{z = 0}$ to $\mathrm{z = 4}$ into 160 bins of size 0.025. The total number of GW events is calculated using the equation 
\begin{equation}
\mathrm{N_{GW}=T_{obs}\int_0^z\frac{dV_c}{dz}\frac{R(z)}{(1+z)}dz},
\label{eq:TotEvent}
\end{equation}
where $\mathrm{\frac{dV_c}{dz}}$ represents the differential comoving volume, $\mathrm{R(z)}$ is the redshift-dependent merger rate, and $\mathrm{T_{obs}}$ is the total observational time in years. The factor $\mathrm{(1+z)}$ accounts for the time dilation effect due to cosmic expansion.

\begin{figure}
    \centering
    \includegraphics[width=8cm, height=6.5cm]{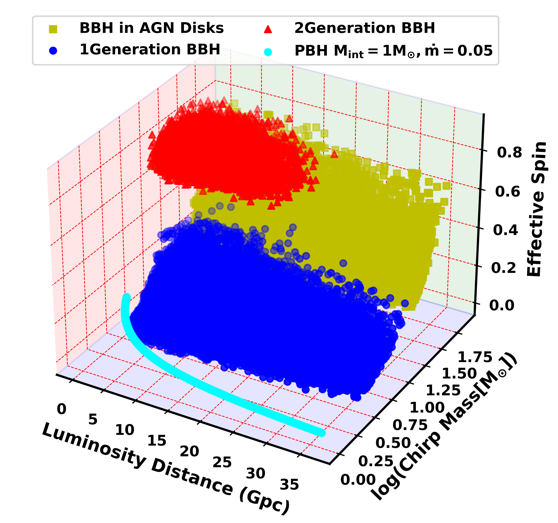}
    \caption{This 3D plot illustrates the phase space distribution of ABHs and PBHs, highlighting the distribution of chirp mass, luminosity distance, and effective spin parameter. It shows the distinct regions in the phase space occupied by 1G black holes, 2G black holes, black holes formed in AGN disks, and PBHs.}
    \label{fig:PhaseSpace}
\end{figure}

For each population, based on its respective merger rate over an observation period of 26 months, we generate GW events by sampling mass and spin values from their distributions using the cumulative distribution function (CDF) method. From these samples, we calculate the chirp mass and effective spin, and then plot the resulting events in phase space, as illustrated in Figure \ref{fig:PhaseSpace}.

For the generation of the phase space (Figure \ref{fig:PhaseSpace}), we have used the following parameter values for mass distribution, which is parameterized by a combination of Gaussian and exponential decay functions  (defined in Section \ref{sec:MassSpinABH}): for 1G black holes, $\mathrm{M_{median} = 8}$ $M_\odot$, $\mathrm{\sigma = 1.5}$ $M_\odot$, and $\mathrm{\alpha= 0.15}$; for 2G black holes, $\mathrm{M_{median} = 16}$ $M_\odot$, $\mathrm{\sigma = 2.0}$ $M_\odot$, and $\mathrm{\alpha= 0.06}$; and for black holes in AGN disks, $\mathrm{M_{median} = 30}$ $M_\odot$, $\mathrm{\sigma = 2.5}$ $M_\odot$, and $\mathrm{\alpha= 0.05}$ to allow for a broader mass spectrum, as shown in Figure \ref{fig:ABHMassModel}. For the spin parameter, due to limited observational knowledge, we parameterize the spin as a function of mass which can capture a wide range of spin behaviors (defined in Section \ref{sec:MassSpinABH}), and is illustrated in Figure \ref{fig:ABHMassCorSpin}. For 1G and AGN disk black holes, we use this parametrization  with the following parameter values: \(m_{\text{min}} = 5M_\odot\), \(m_{\text{max}} = 120M_\odot\), \(n = 1.0\), \(\beta = 0.2\), and \(\gamma = 0.01\). For 2G black holes, we assume a Gaussian distribution centered at a spin of 0.7 with a standard deviation of 0.1. In our study, we assume a delay time of 500 Myrs for 1G black holes and 1 Gyrs for 2G black holes. For the merger rate of BBHs in AGN disks, we utilize the description provided in \cite{Yang:2020lhq}, which is detailed in Appendix \ref{sec:AGNMerg}.

Additionally, for PBHs, we select an initial mass of \(1\,M_\odot\) and set the mass accretion index to \(\dot{m} = 0.05\). The corresponding merger rate is calculated using a PBH fraction of \(f_{\rm pbh} = 0.001\), as described in Section \ref{sec:PBHDetails}.

For the mass distributions, we adopt log-normal distributions (defined in Section \ref{sec:PBHDetails}), with the characteristic mass evolving with redshift according to Equation \eqref{eq:MassEvoPBH}. The standard deviation is set to 2\% of the characteristic mass, providing a baseline for exploring the parameter space without introducing strong biases. Similarly, for the spin distribution, we use the median value defined in Equation \eqref{eq:SpinEvoPBH}, with an added 2\% Gaussian noise. From these sampled values, we calculate the chirp mass and effective spin, and plot the resulting events in phase space (Figure \ref{fig:PhaseSpace}). Although realistic noise levels may be higher, our analysis shows that increasing the noise level has only a minor impact on our results, primarily because very few data points lie along the PBH trajectories.

We have varied some parameters that have the most significant impact on the phase space to better analyze and infer information about the formation channels (Figure \ref{fig:ProjectionResult}). Figure~\ref{fig:PhaseSpace} presents a visualization of the phase space occupied by black holes from four different formation channels, plotted in terms of chirp mass, luminosity distance, and effective spin. The figure clearly demonstrates that these populations occupy distinct regions in the phase space, although some overlap may occur particularly between black holes in AGN disks and 2G black holes, as well as between AGN black holes and 1G black holes. 

This degeneracy can be resolved by introducing a fourth parameter: eccentricity. Isolated binary evolution generally results in nearly circular orbits due to mass loss from stellar winds and GW emission \citep{Hurley:2002rf}. In contrast, binaries formed through dynamical interactions can retain higher eccentricities, especially during close encounters. Hierarchical mergers can also yield eccentric orbits depending on the dynamics involved \citep{Antonini:2016gqe}. This can be included in phases-space for future GW analysis.

\vspace{0.25cm}
\noindent\textbf{Impact of Metallicity Evolution on Phase Space}
\vspace{0.15cm}

Another crucial factor that can influence the location and spread of black hole populations in phase space is the metallicity of progenitor stars. However, since most of the GW events analyzed in this study lie at relatively low redshifts, we do not include metallicity evolution in our main analysis. Nevertheless, our phase-space reconstruction technique is capable of capturing such effects, and we demonstrate this sensitivity by modeling how metallicity evolution can influence the 1G black hole population.

\begin{figure*}
\includegraphics[width=\textwidth, height=5.6cm]{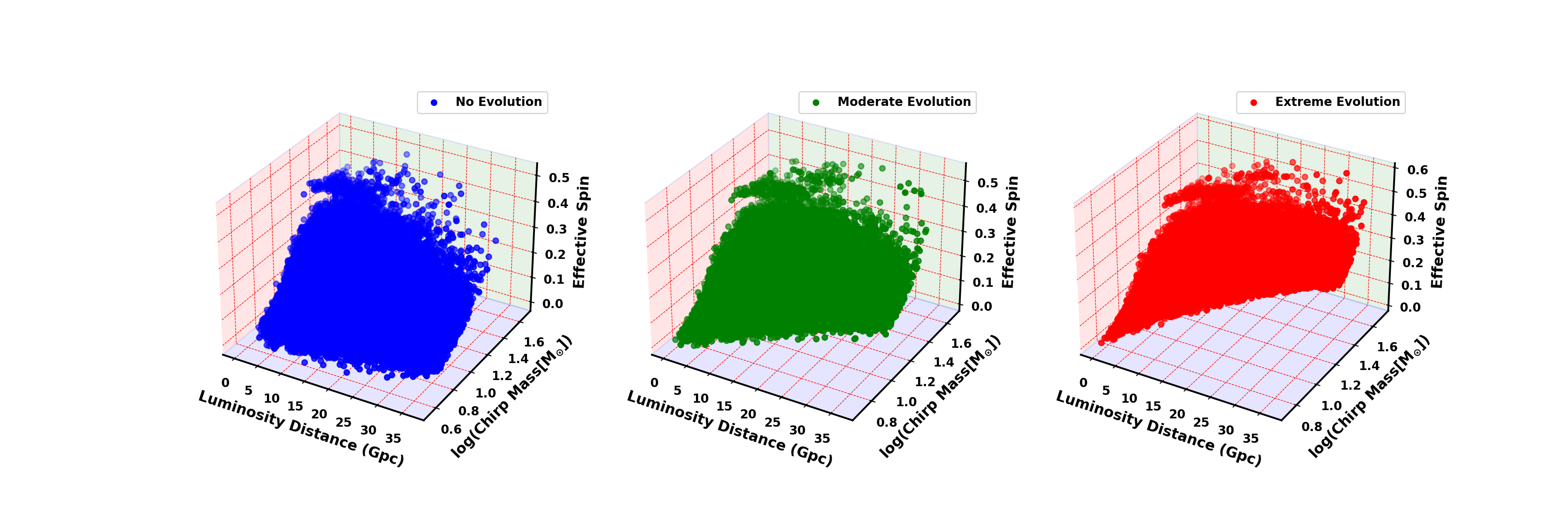}
\caption{Phase space distribution of 1G black holes illustrating the impact of metallicity evolution. The three panels represent different scenarios: (left) no evolution (\(\delta = 0\)), (middle) moderate evolution (\(\delta = 0.25\)), and (right) extreme evolution (\(\delta = 1\)). As metallicity evolution increases, the overall mass distribution shifts towards higher values with redshift, and the lower boundary of the distribution systematically rises with \(\delta\). These distinct signatures demonstrate the sensitivity of our phase-space approach in detecting metallicity-driven mass evolution.}
\label{fig:PhaseSpace2}
\end{figure*}

In Figure \ref{fig:PhaseSpace2}, we illustrate the impact of metallicity evolution on the phase space for 1G black holes. To model this effect, we assume a linear dependence on redshift, given by \cite{Mukherjee:2021rtw, Karathanasis:2022rtr}
\begin{equation}
    \mathrm{M_{\text{median}}(z) = M_{\text{median}}(1 + \delta\times z)}.
    \label{eq:MetEvo}
\end{equation}
While the specific features may vary depending on the modeling approach, the overall trends remain distinct and observable in the phase space \citep{Karathanasis:2022rtr}. It is important to point out that this is a simple extension to capture the redshift evolution of metallicity and its impact on GW sources. But such simple models are also difficult to be constrained from the current data \cite{Karathanasis:2022rtr}. In the future, a more complex model can be considered and tested from future GW observations. Currently, we consider three scenarios: (i) no evolution, where the median mass remains constant at \( M_{\text{median}} = 8M_{\odot} \), corresponding to \( \delta = 0 \); (ii) moderate evolution, with \( \delta = 0.25 \); and (iii) extreme evolution, where \( \delta = 1 \). 

From Figure \ref{fig:PhaseSpace2}, it is evident that as redshift increases, the overall mass distribution shifts towards higher values, and the lower boundary of the distribution exhibits a systematic increase proportional to \(\delta\), resulting in a more pronounced separation in the phase space. This trend is particularly noticeable in the extreme evolution case (\(\delta = 1\)), where the mass distribution extends significantly towards higher redshifts. These distinct signatures highlight the sensitivity of our phase-space approach in detecting and characterizing metallicity-driven mass evolution.

Figure \ref{fig:PhaseSpace2} is constructed using the same methodology described earlier for Figure \ref{fig:PhaseSpace}, with the only modification being the inclusion of metallicity evolution through a redshift-dependent shift in the median mass as defined by Equation \ref{eq:MetEvo}. All other parameters and sampling methods remain unchanged.

\section{Classification of Compact Binary Events by Formation Channel}
\label{sec:classification}

We investigate the classification of GW events based on their mass characteristics. This classification is crucial for understanding the potential origins of these events, particularly in distinguishing between different BBH formation channels \citep{Mould:2023ift,Kimball:2019mfs,Andrews:2020pjg}. As illustrated in Figure \ref{fig:GWCatelog}, we categorize each compact binary event from the GWTC-3 catalog, including GW230529, while excluding GW170817, which has been confirmed as a binary neutron star (BNS) system due to its observed electromagnetic counterpart \citep{LIGOScientific:2017vwq}.

In Figure \ref{fig:GWCatelog}, each pie chart depicts the probabilities of each event originating from 1G BBH formation channels or a combined category of 2G BBHs and BBHs formed in AGN disks (denoted as 2G+AGN). The top right portion of each pie chart represents the probability of 1G origins, while the left portion indicates the probability of 2G+AGN origins. These probabilities are not normalized per event and therefore are not expected to sum to unity. Instead, they are derived from channel-specific probability densities across the entire phase space and should be interpreted in that broader context. Events that exhibit a significant probability contribution from either the 1G or the combined 2G+AGN categories are marked in black, indicating a stronger probability of belonging to these channels. In contrast, events with negligible contributions are highlighted in gray indicating them as \textit{confusing sources}, suggesting they may be potential PBH candidates or channels associated with binary neutron stars which are not considered in this analysis. The overlaid numbers for these events represent the $\mathrm{M_{int}}$ and $\mathrm{\dot{m}}$ values for which the event could be considered a candidate for PBH. Note that these values represent only one possible combination; many other combinations are also possible.

To calculate the probability that a BCO event originates from different formation channels, we model the mass distribution for three distinct black hole populations: 1G black holes, 2G black holes, and black holes in AGN disks using a power-law plus Gaussian component as detailed in Equation \ref{eq:MassModel}, with distinct parameter values for each formation channel.

For 1G black holes, the mass distribution is characterized by the following parameters: $\mathrm{M_{median} = 8} \, M_\odot, \, \sigma = 1.5 \, M_\odot, \, \alpha = 0.15.$ For 2G black holes, the parameters are: $\mathrm{M_{median} = 16} \, M_\odot, \, \sigma = 2.0 \, M_\odot, \, \alpha = 0.06.$ For black holes in AGN disks, the mass distribution is given by: $\mathrm{M_{median} = 30} \, M_\odot, \, \sigma = 2.5 \, M_\odot, \, \alpha = 0.05.$ The probability distribution of the chirp mass \( P(M_{\text{chirp}}) \) is then computed by marginalizing over the mass distributions of the two individual black holes in the binary system. The chirp mass \( M_c \) is related to the component masses \( m_1 \) and \( m_2 \) by:

\begin{equation}
    \mathrm{M_c} = \frac{(m_1 m_2)^{3/5}}{(m_1 + m_2)^{1/5}}
\end{equation}

To calculate the chirp mass distribution \( P(M_c) \), we integrate over the mass distributions \( P(m_1) \) and \( P(m_2) \) of the two component black holes:

\begin{equation}
\begin{split}
    \mathrm{P(M_c)} = & \int \int P(m_1) P(m_2) \\
    & \times \delta\left(M_c - \frac{(m_1 m_2)^{3/5}}{(m_1 + m_2)^{1/5}}\right) \, dm_1 \, dm_2
\end{split}
\end{equation}

where \( \delta \) is the Dirac delta function, ensuring that only the correct combinations of \( m_1 \) and \( m_2 \) contribute to the chirp mass \( M_c \). Finally, to compute the probability that an observed binary event originates from a particular formation channel, we use the following expression for the probability \( w_{\text{channel}} \):

\begin{equation}
    \mathrm{w_{channel} = \frac{\int P_{event}(M_c) P_{channel}(M_{chirp}) \, dM_{chirp}}{\int P_{\text{channel}}(M_{chirp}) \, dM_{chirp}}.}
\end{equation}

Here, \( P_{\text{event}}(M_{\text{chirp}}) \) represents the likelihood of chirp mass of the observed event, and \( P_{\text{channel}}(M_{\text{chirp}}) \) represents the mass distribution for a specific formation channel. The integration is performed over the entire range of chirp masses \( M_{\text{chirp}} \).

In the catalog, there are six events (GW190425, GW190814, GW190917, GW191219, GW200115, and GW230529) that are unlikely to originate from standard BBH formation channels, as they exhibit relatively lower masses. Additionally, the overlaid numbers on the pie charts correspond to the values of \(\mathrm{M_{int}}\) and \(\mathrm{\dot{m}}\) for these events, indicating the conditions under which they could be considered PBH candidates. While the probability of an event originating from a particular formation channel may shift with changes in the parameters governing the formation channels, it is unlikely that the classification of these six events would be significantly affected. These events are positioned at the lower end of the mass spectrum, making it improbable that they originate from any standard astrophysical formation channel. Thus, even with parameter variations, their classification remains robust.

\section{Finding Black Hole formation channels from the GWTC-3 Catalog}
\label{sec:Result}

In this study, we analyze the phase space of observed GW events using the GWTC-3 catalog \citep{KAGRA:2021duu}, along with the publicly available event GW230529 \citep{LIGOScientific:2024elc}. The compact object phase space is generally defined in an $N$-dimensional observable space, where each dimension corresponds to a measurable parameter inferred from GW observations. However, for this analysis, we restrict our focus to a three-dimensional subspace spanned by the chirp mass $\mathrm{M_c}$, the effective spin $\mathrm{\chi_{eff}}$, and the luminosity distance $\mathrm{D_L}$.

We construct the observed phase space directly from the posterior probability distributions provided by gravitational-wave parameter estimation methods. Each GW event yields a posterior distribution \( P_{\text{GW}}(M_c, \chi_{\mathrm{eff}}, D_L) \), which encapsulates the measurement uncertainties and correlations among these parameters. These distributions are individually normalized to obtain the event-specific phase space density \( Z_i(M_c, \chi_{\mathrm{eff}}, D_L) \). By summing the normalized densities across all events, we form the cumulative observational phase space \( Z_{\text{total}}(M_c, \chi_{\mathrm{eff}}, D_L) \), which naturally incorporates measurement uncertainties and reflects the full statistical structure of the observed population.

\begin{figure*}
\includegraphics[width=0.24\textwidth, height=4.0cm]{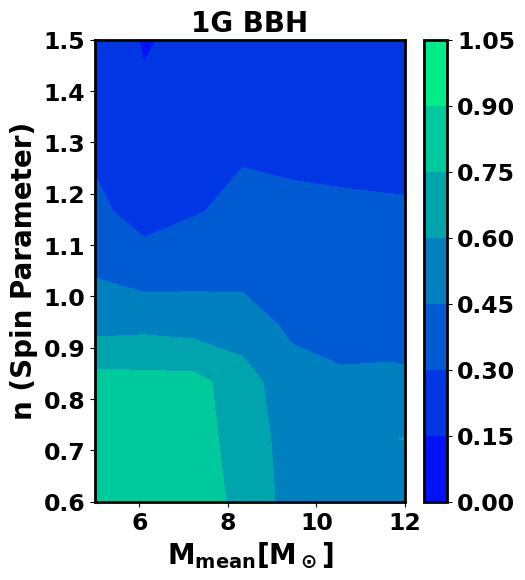}
\includegraphics[width=0.24\textwidth, height=4.0cm]{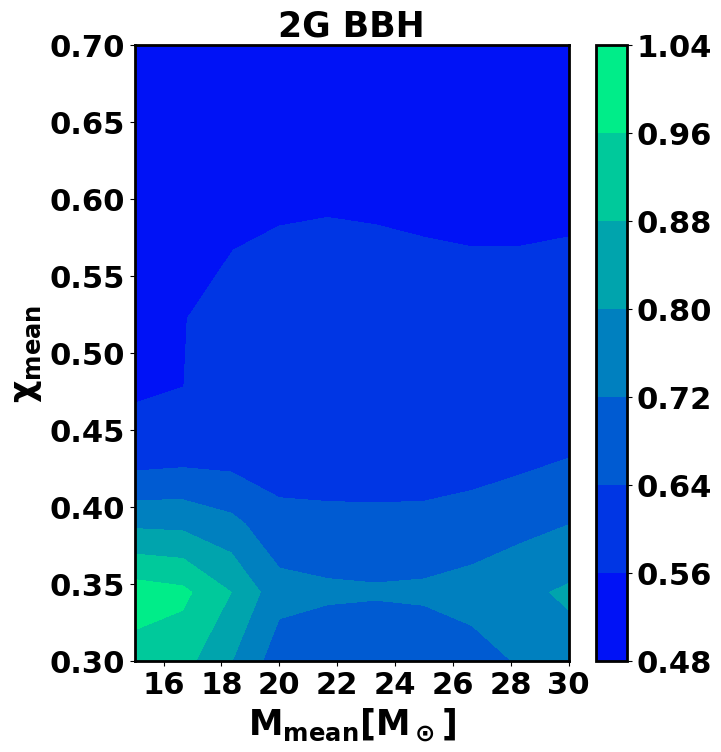}
\includegraphics[width=0.24\textwidth, height=4.0cm]{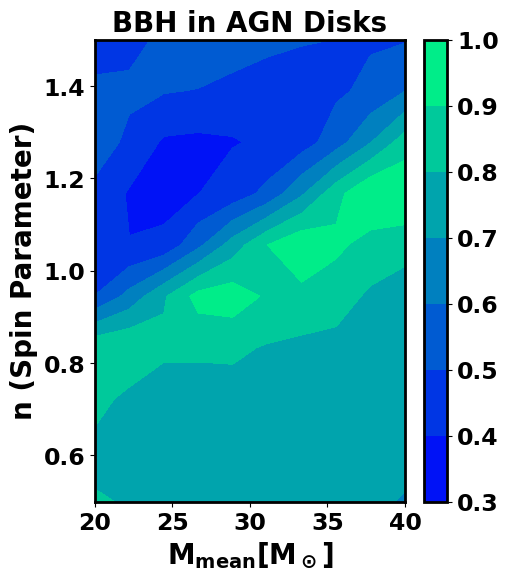}
\includegraphics[width=0.24\textwidth, height=4.0cm]{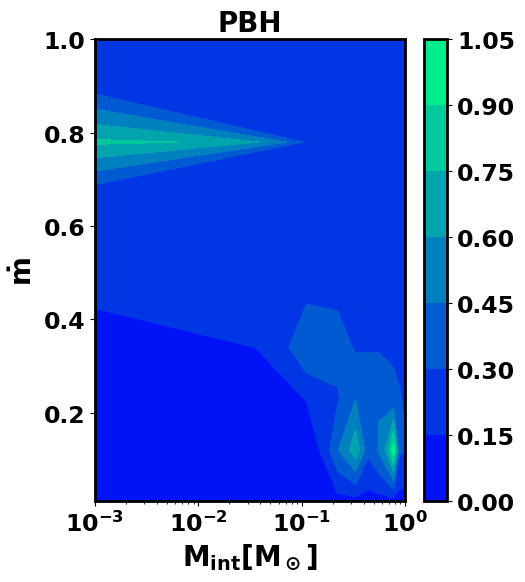}

\caption{Projection of phase space trajectories for different BBH formation channels. Leftmost: 1G BBHs, illustrating the probability of phase space trajectories generated by variations in the spin parameter $\mathrm{n}$ and the median mass $\mathrm{M_{median}}$. Second from the left: 2G BBHs, highlighting the probability associated with variations in the median mass $\mathrm{M_{median}}$ and the median spin parameter $\mathrm{\chi_{median}}$, assuming a Gaussian distribution for the spin. Third from the left: Binary black holes formed in AGN disks, depicting the probability of trajectories influenced by variations in the spin parameter $\mathrm{n}$ and the median mass $\mathrm{M_{median}}$. Rightmost: PBHs, showcasing the probability of their trajectories within the phase space based on the mass accretion rate index $\mathrm{\dot{m}}$ and the initial mass $\mathrm{M_{int}}$. The color bar in all plots represents the probability value.}
\label{fig:ProjectionResult}
\end{figure*}

To explore the theoretical trajectories corresponding to different formation channels, we focus on two key model parameters that significantly influence the shape of the phase space trajectories for both PBHs and ABHs. For PBHs, these parameters are the mass accretion rate index, $\mathrm{\dot{m}}$, and the initial mass, $\mathrm{M_{int}}$. In the case of ABHs, we consider three distinct formation channels: 1G BBHs, 2G BBHs, and BBHs forming in AGN disks. For 1G BBHs and those formed in AGN disks, we vary the spin parameter denoted by $\mathrm{n}$ and the median of the Gaussian mass distribution denoted by $\mathrm{M_{median}}$. For 2G BBHs, we explore both the median mass, $\mathrm{M_{median}}$, and the median spin parameter, $\mathrm{\chi_{median}}$, as the spin distribution for 2G BBHs is assumed to follow a Gaussian distribution (See Section \ref{sec:MassSpinABH} for more details.). By systematically varying key parameters for each of the four cases: PBHs, 1G BBHs, 2G BBHs, and AGN disk BBHs we explore a wide range of the phase space.

For 1G BBHs, we vary the median mass ($M_{\mathrm{median}}$) from 5 to 12 $\mathrm{M_{\odot}}$, for 2G BBHs from 15 to 30 $\mathrm{M_{\odot}}$, and for BBHs in AGN disks from 20 to 40 $\mathrm{M_{\odot}}$. The spin parameter ($n$) is varied from 0.6 to 1.5 for 1G BBHs, and from 0.5 to 1.5 for BBHs in AGN disks. We also vary the median spin ($\chi_{\mathrm{median}}$) from 0.30 to 0.70 for 2G BBHs. Importantly, our parametric framework is general enough to capture the dynamical effects in cluster channels where BBHs show broader mass distributions due to hierarchical mergers and dynamical mass segregation, along with more randomized spins by allowing wide variations in mass and spin parameters  \citep{Fragione:2023kqv,Rodriguez:2016,Zevin:2021,Mapelli:2020vfa,Antonini:2020xnd}.

For PBHs, the initial mass ($M_{\mathrm{int}}$) ranges from $10^{-3}$ to 1 $\mathrm{M_{\odot}}$, which is sufficiently high to avoid significant evaporation due to Hawking radiation over the age of the Universe \citep{Carr:2009jm}, and accretion rate ($\dot{m}$) spans from 0 to 1, covering sub-Eddington to Eddington limits. While alternative accretion scenarios or spin evolution might affect the dynamics, current GW data are not informative enough to infer a more complex model. Thus, we restrict our analysis to simple Eddington accretion model. We fix the spin evolution parameter $k$ for PBHs to 10, and for ABH spin cases, we set $\beta = 0.8$ and $\gamma = 0.04$ due to their minimal impact on trajectories. This broad parameter space serves as a prior in the likelihood analysis, enabling the generation of diverse phase space trajectories. We adopt this parametric form because the current GW dataset is insufficient, with only a few tens of high-SNR events available to resolve finer astrophysical details (e.g., metallicity, reaction rate uncertainties, etc.), yet our phase-space approach can capture any observable complexities in the model.

We project the theoretically generated trajectories for different formation channels onto the observed \texttt{BCO Phase Space}, which was constructed from GWTC-3 events and GW230529, as described above. This projection enables a direct comparison between model predictions and observational data within the common parameter space. To ensure that only detectable events are considered in our comparison, we apply a selection function that effectively filters out parameter combinations corresponding to sources that would not be detected by the LVK detectors. In this study, the selection function is constructed based on the mass and luminosity distance of the binary system-two parameters that significantly affect the signal-to-noise ratio.

For each combination of model parameters across the four formation scenarios\text{-}1G BBHs, 2G BBHs, BBHs in AGN disks, and PBHs\text{-}we generate a set of allowed trajectories in the observable phase space, specifically defined by the three parameters \((M_c, \chi_{\mathrm{eff}}, D_L)\). These trajectories are constructed based on the underlying model assumptions and varied over the corresponding parameter grids. After applying the selection function to exclude undetectable points, we evaluate the overlap between these model trajectories and the observed phase space density. This is quantified by computing the total probability support from observations for each set of model parameters:

\begin{equation}
\mathrm{P_{\text{trajectories}}} = \sum_{(M_c, \chi_{\mathrm{eff}}, D_L) \in \text{trajectories}} Z_{\text{total}}(M_c, \chi_{\mathrm{eff}}, D_L),
\end{equation}

where \(\mathrm{P_{\text{trajectories}}}\) represents the cumulative probability assigned to a trajectory by the observed compact object phase space. The summation runs over all phase space points \((M_c, \chi_{\mathrm{eff}}, D_L)\) along the theoretical trajectory that satisfy the detectability condition enforced by the selection function. The term \(Z_{\text{total}}(M_c, \chi_{\mathrm{eff}}, D_L)\) corresponds to the value of the total observed phase space density at each of these detectable model points. This process allows us to assess how well different formation channels explain the observed population, while accounting for both detector selection effects and measurement uncertainties.

The probability results for the four scenarios are presented in Figure \ref{fig:ProjectionResult}. For 1G BBH, these probabilities suggest that it is more likely to have a low median mass and a shallow spin growth $\mathrm{n} < 1$ value. This arises as most of the sources in the GWTC-3 indicate low mass and non-spinning. In contrast, for BBHs in AGN disks, a higher median mass M$_{\rm median}$ and slightly higher values of $\mathrm{n}$ are more probable, suggesting there can be significant growth of spin. This arises due to a few GW events with heavier mass and non-zero spin. For the 2G BBH formation channel, the results are less conclusive, as the phase space regions that exhibit higher probabilities can show significant degeneracy with other channels. However, if we ignore this degeneracy, slightly higher median masses (around M$_{\rm median}= 16$ M$_\odot$) with a median spin around $\chi_{\rm median}= 0.35$ appear to be somewhat more probable than other combinations. 

In the case of PBHs, the probability is generally very low across most regions in the parameter space. However, certain regions offer some support for PBHs. One such region is for PBHs with masses close to 0.1 to 1 $\mathrm{M_\odot}$ and low accretion rates (around $\mathrm{\dot{m}} \sim 0.1$). This is supported by observations of low-redshift, low-mass, and high-spin events. Additionally, another region that provides support involves PBHs with initial masses between 0.001 and 0.01 $\mathrm{M_\odot}$ and high accretion rates (around $\mathrm{\dot{m}} \sim 0.8$), which align with events characterized by chirp masses of $\sim$ 30 $\mathrm{M_\odot}$ at redshifts around 0.9. This low support is arising because the PBH trajectory differs from the ABH trajectory, but at higher masses (as shown in Figure \ref{fig:PhaseSpace}), there can be degeneracy with 2G and AGN black holes at high masses. Among these, GW190425 and GW230529 are particularly notable due to their significantly high spins, which matches with a possible hypothesis that these have undergone accretion, making them possible candidates for further investigation as PBHs. To further explore the potential PBH origin of these events, we provide the specific combinations of initial masses $\mathrm{M_{int}}$,  accretion rates $\mathrm{\dot{m}}$, and spin parameters $\mathrm{k}$ in Table \ref{tab:PBHParam} for which each event individually falls along a PBH trajectory in the \texttt{BCO Phase Space} within the $68\%$ C.I. of their respective posteriors. It is important to note these constraints are completely independent of any constraints feasible from the stochastic GW background \citep{Wang:2016ana,Cai:2021wzd,Mandic:2016lcn}.

An alternative astrophysical explanation for these events could be the formation of second-generation BBHs from the merger of two neutron stars, heavy neutron stars, or any exotic compact objects, or this may be viewed in light of the uncertainties regarding core-collapse supernovae and fallback efficiency \citep{2020ApJ...890...51E,Chan:2020lnd}, as well as the systematics inherent in the GW side \citep{LIGOScientific:2016ebw,Pankow:2016udj,Sampson:2013jpa}. In the future, exploration of the \texttt{BCO Phase Space} for neutron stars (or exotic compact objects) along with PBHs will shed more light on such possibilities with the availability of more events by identifying which of these trajectories is getting more occupied with GW sources.

\begin{table}
\centering
\begin{tabular}{|l|c|c|c|c|c|c|}
\hline
Parameters & Case-1 & Case-2 & Case-3 & Case-4 & Case-5 & Case-6 \\
\hline
$\mathrm{M_{int}}$ & 0.0100 & 0.0703 & 0.1306 & 0.1708 & 0.2311 & 0.3115 \\
\hline
$\mathrm{\dot{m}}$ & 0.18 & 0.11 & 0.09 & 0.08 & 0.07 & 0.06  \\
\hline
$\mathrm{k}$ & 7 & 8 & 9 & 10 & 10 & 11  \\
\hline
\end{tabular}
\caption{The table lists six combinations of initial PBH masses ($M_{\rm int}$), accretion rates $\dot m$, and spin parameters ($\mathrm{k}$), each corresponding to a distinct trajectory in the PBH phase space that is consistent with both the GW230529 and GW190429 events within the $68\%$ C.I.}
\label{tab:PBHParam}
\end{table}

\textbf{\textit{Assumption and the main caveats:}} We note that while our current analysis employs simplified parametric model for both astrophysical and primordial formation channels for applying to the latest catalog, which does not capture the full complexity of detailed astrophysical modeling (such as fallback physics, supernova explosion models, mass-loss prescriptions, and the evolution of metallicity) our \texttt{BCO Phase Space} technique is fundamentally designed to be flexible and extendable to any model. This simplified approach was chosen because the current data is not yet informative enough to robustly constrain more complex properties, even though our phase-space methodology is fully capable of capturing such effects, we cannot distinguish between different models at present. Also, the identification of PBH-related events and black holes in AGN distributions remains preliminary and uncertain, and in the future more data with high-SNR GW observations will allow us to incorporate deeper insight into these channels and also use more advanced modeling of cluster dynamics, fallback physics, and metallicity evolution, thereby refining parameter constraints and enhancing discrimination between astrophysical and primordial channels.

\section{Conclusions} 
\label{sec:summary}
In this work, we introduced a novel framework for identifying the formation channels of BCOs such as neutron stars and black holes by leveraging the phase space geometry of GW events. Our method offers a flexible, data-driven approach by directly analyzing the phase space trajectories that represent different BCO formation channels. Each formation channel has distinct characteristics, placing it in a unique region of the phase space. By projecting these theoretical trajectories onto the observed phase space, we demonstrate that it is possible to infer key information related to the formation channels of the BCOs detected.   

By applying this technique to the GWTC-3 catalog and GW230529 we explore different physical scenarios of formation associated with these binaries. Our findings show that most of the GW events are classified in 
first-generation and second-generation scenarios, except six events that are in the mass gap (as shown in Figure \ref{fig:GWCatelog}. A closure inspection reveals that two events GW190425 and GW230529 can be associated with a common origin of a sub-solar black hole which has gone through a similar sub-Eddington accretion to grow into a mass and spin with which it is detected by the GW detectors. Though this finding from \texttt{BCO Phase Space} analysis is only an initial hint, other possible astrophysical scenarios of low mass-gap events are yet to be explored. However in the future with more detected GW events, it will be possible to distinguish whether more low-mass events are appearing on the PBH trajectory or ABH trajectory. Either way, this new revelation using \texttt{BCO Phase Space} will make a paradigm shift in our understanding of the formation and evolution of compact objects.

As \texttt{BCO Phase Space} can capture any scenarios of formation channel (known or unknown), it will be possible  to make serendipitous discoveries of new formation scenarios of either astrophysical or primordial origin using this technique. With the future LVK observations, this \texttt{BCO Phase Space} technique will be extended for neutron stars, and globular clusters formation scenarios and will also be extended for its application on future GW detectors such as Cosmic Explorer \citep{Reitze:2019iox}, Einstein Telescope \citep{Maggiore:2019uih}, and LISA \citep{Colpi:2024xhw}, along with the inclusion of other observables such as eccentricity and kick velocity. These advanced detectors, with their ability to measure wider frequency ranges, will provide significantly improved accuracy that will enable more precise characterization of binary formation channels and will bring deeper insights into their formation and cosmic evolution.

\section*{Acknowledgments}
The authors express their gratitude to Michela Mapelli for reviewing the manuscript and providing useful comments as a part of the LIGO publication policy. This work is part of the \texttt{⟨data|theory⟩ Universe-Lab}, supported by TIFR and the Department of Atomic Energy, Government of India. The authors express gratitude to the system administrator of the computer cluster of \texttt{⟨data|theory⟩ Universe-Lab} and the TIFR computer center HPC facility for computing resources. Special thanks to the LIGO-Virgo-KAGRA Scientific Collaboration for providing noise curves. LIGO, funded by the U.S. National Science Foundation (NSF), and Virgo, supported by the French CNRS, Italian INFN, and Dutch Nikhef, along with contributions from Polish and Hungarian institutes. This collaborative effort is backed by the NSF’s LIGO Laboratory, a major facility fully funded by the National Science Foundation. The research leverages data and software from the Gravitational Wave Open Science Center, a service provided by LIGO Laboratory, the LIGO Scientific Collaboration, Virgo Collaboration, and KAGRA. Advanced LIGO's construction and operation receive support from STFC of the UK, Max-Planck Society (MPS), and the State of Niedersachsen/Germany, with additional backing from the Australian Research Council. Virgo, affiliated with the European Gravitational Observatory (EGO), secures funding through contributions from various European institutions. Meanwhile, KAGRA's construction and operation are funded by MEXT, JSPS, NRF, MSIT, AS, and MoST. This material is based upon work supported by NSF’s LIGO Laboratory which is a major facility fully funded by the National Science Foundation. We acknowledge the use of the following packages in this work: Numpy \citep{van2011numpy}, Scipy \citep{jones2001scipy}, Matplotlib \citep{hunter2007matplotlib}, and Astropy \citep{robitaille2013astropy}.

\appendix
\section{Merger Rate of black holes in AGN Disks}
\label{sec:AGNMerg}

The active galactic nucleus (AGN) disk environment offers a unique setting where black hole mergers can occur at an elevated rate compared to other astrophysical scenarios. The dense gas in AGN disks fosters both black hole growth through accretion and dynamical interactions that lead to frequent mergers. In such environments, the high density of stellar-mass black holes within the disk promotes the formation of binaries that can merge due to gas torques, gravitational wave emission, and multi-body interactions. The presence of this gas-rich medium not only accelerates black hole mass growth via accretion but also increases the probability of repeated mergers, especially among lower-mass black holes that can evolve into more massive systems\citep{Tagawa:2019osr,2024arXiv240216948F, Wang:2021clu}.

The rate of black hole mergers in AGN disks is significantly enhanced because the gas drives black holes toward the disk midplane, where they become trapped and accumulate in large numbers. As black holes interact within this high-density region, the probability of binary formation and subsequent mergers increases. A key factor determining the black hole merger rate in AGN disks is the population density of AGNs, which evolves with redshift. This density can be constrained using the AGN luminosity function (LF), $f_L(L, z)$, which describes the distribution of AGNs as a function of their luminosity and redshift. The bolometric AGN LF provides insights into the overall energy output of AGNs, which is related to the number of black holes in their disks capable of undergoing mergers, can be expressed as \citep{Shen:2020obl}:

\begin{equation}
    f_L(L, z) = \frac{f_*(L)}{\left(\frac{L}{L_*}\right)^{\gamma_1(z)} + \left(\frac{L}{L_*}\right)^{\gamma_2(z)}},
\end{equation}

where $\gamma_1(z)$, $\gamma_2(z)$, and $L_*(z)$ are given by the following relations:

\begin{equation}
    \gamma_1(z) = \alpha_0T_0(1+z) + \alpha_1 T_1(1+z) + \alpha_2 T_2(1+z),
\end{equation}

\begin{equation}
    \gamma_2(z) = \frac{2b_0}{\left(\frac{1+z}{3}\right)^{b_1} + \left(\frac{1+z}{3}\right)^{b_2}},
\end{equation}

\begin{equation}
    \log L_*(z) = \frac{2c_0}{\left(\frac{1+z}{3}\right)^{c_1} + \left(\frac{1+z}{3}\right)^{c_2}},
\end{equation}

\begin{equation}
    \log f_*(z) = d_0T_0(1+z) + d_1 T_1(1+z),
\end{equation}

where $T_n(z)$ are Chebyshev polynomials. The best-fit parameters are listed in the Table \ref{tab:AGNparam}.

\begin{table*}
\centering
\begin{tabular}{|l|c|c|c|c|c|c|c|c|c|c|}
\hline
$\alpha_0$ & $\alpha_1$ & $\alpha_2$ & $\beta_0$ & $\beta_1$ & $\beta_2$ & $c_0$ & $c_1$ & $c_2$ & $d_0$ & $d_1$\\
\hline
0.8396     & -0.2519    & 0.0198     & 2.5432     & -1.0528    & 1.1284     & 13.0124 & -0.5777 & 0.4545 & -3.5148 & -0.4045 \\
\hline
\end{tabular}
\caption{Best Fit Parameter Values for the AGN Luminosity Function}
\label{tab:AGNparam}
\end{table*}

The AGN density $n_{\rm AGN}(z)$ can be obtained through direct integration of $f_L(L, z)$ with a lower cutoff $L_{\rm min}$ due to large uncertainties. The mass of supermassive black holes (SMBHs) correlates with AGN luminosity via:

\begin{equation}
    \frac{M_\bullet}{M_\odot} = 3.17 \times 10^{-5}\frac{1-\epsilon}{\dot{m}}\frac{L}{L_\odot},
\end{equation}
where $\epsilon$ is the radiation efficiency and $\dot{M}$ is the SMBH accretion rate. This relationship can be reformulated to link the normalized accretion rate $\dot{m}$ with the Eddington ratio $\lambda = L/L_{\rm Edd}$. The Eddington ratio distribution is expressed as a mixture model \citep{Tucci:2016tyc}:
\begin{equation}
    P(\lambda | L, z) = f_{\rm uno} P_1(\lambda | z) + f_{\rm obs} P_2(\lambda | z),
\end{equation}
where $f_{\rm uno} = 1 - f_{\rm obs}$ is the fraction of unobscured (type-1) AGNs and $f_{\rm obs}$ the fraction of obscured (type-2) AGNs. For type-1 AGNs, $P_1(\lambda | z)$ follows a log-normal distribution with:
\begin{align}
    \ln \lambda_{\rm c}(z) &= \max\left[1.9 - 0.45z, \ln 0.03\right], \\
    \sigma_z &= \max\left[1.03 - 0.15z, 0.6\right].
\end{align}
For type-2 AGNs, $P_2(\lambda | z)$ follows a gamma distribution with a low-Eddington cutoff 
\begin{equation}
    P_2(\lambda | z) = N_2(z) \lambda^{\alpha(z)} \exp(-\lambda / \lambda_0),
\end{equation}
where $\lambda_0 = 1.5$, and $\alpha(z)$ is defined as $0.6$ for $z < 0.6$ and $0.4 + 0.6(z - 0.6)$ for $z \geq 0.6$. The fraction of obscured AGNs, $f_{\rm obs}$, is parameterized as \citep{Ueda:2014tma}
\begin{equation}
    f_{\rm obs} = \frac{(1 + f_{\rm CTK})\psi(L_X, z)}{1 + f_{\rm CTK}\psi(L_X, z)},
\end{equation}
where $f_{\rm CTK} = 1$ is the ratio of Compton-thick to Compton-thin AGNs, and $\psi(L_X, z)$ is given by 
\begin{equation}
    \psi(L_X, z) = \min(\psi_{\rm max}, \beta (\log L_X - 43.75\xi(z)) + \psi_{\rm min}),
\end{equation}
with $\psi_{\rm max} = 0.84$, $\psi_{\rm min} = 0.2$, $\beta = 0.24$, and $\xi(z) = 0.43(1+z)^{0.48}$ \citep{Marconi:2003tg}. The black hole merger rate $\Gamma$ in a single AGN disk depends on the number of stellar black holes in the disk, $N_{\rm disk}$, which follows 
\begin{equation}
    N_{\rm disk}(n_i) = 5.5 n_i^{1/3},
\end{equation}
where $n_i$ is the dominant factor. The average BH merger rate is 
\begin{equation}
    \Gamma(n_i) = \left(1 - e^{-N_{\rm disk}(n_i)} \right)/T_{\rm AGN},
\end{equation}
where $T_{\rm AGN} = 10^7$ years is the AGN lifetime \citep{Yang:2019cbr}. Assuming that black holes in AGN disks hierarchically merge in migration traps, the BH merger rate becomes 
\begin{equation}
    \Gamma(\dot{m}) = N_{\rm disk}(\dot{m}) \frac{1}{T_{\rm AGN}} (1 - e^{-N_{\rm disk}(\dot{m})}).
\end{equation}

Combining these factors, the cosmic black hole merger rate in AGNs as a function of redshift is given by:
\begin{equation}
    R_{\rm AGN}(z) = \int_{L_{\rm min}}^{L_{\rm max}} f_L(L, z) \, d\log L \int_{\lambda} P(\lambda | L, z) \, \Gamma(\dot{m}) \, d\lambda.
\end{equation}
We use the above expression for calculating the BBH mergers rate redshift evolution hosted in AGNs.

\bibliography{references}
\end{document}